
\baselineskip 1cm
\font\vlbf=cmbx12 scaled 1200
\font\lbf=cmbx12
\hfill {SUTDP/93/71/3}

\hfill {February 93/ Bahman 71}

\vskip 2cm

\centerline{\vlbf Exact Solution for the Most General Minimally Coupled}
\centerline {{\vlbf One Dimensional Lattice Gauge Theories}
\footnote*{Supported by Sharif University of Technology}}

\vskip 2cm

\centerline{\bf M. Khorrami}

\centerline{\it Department of Physics, Sharif University of Technology}

\centerline{\it P.O.Box 11365-9161, Tehran, Iran}

\vskip 2cm

\centerline {\lbf Abstract }

\noindent We consider one dimensional lattice gauge theories constructed by
the minimal coupling prescription. It is shown that these theories are
exactly solvable in the thermodynamic limit. After considering the most
general case, we discuss some special cases on finite lattices, and also work
out some examples. There is no phase transition in these minimally coupled
theories.
\vfil
\noindent{\lbf O Introduction}

\noindent During the last decades lattice gauge theories have been
extensively studied [1-4]. Lattice theories have no ultraviolet divergences,
they provide a non-perturbative approach to some theories, such as QCD [1]
(and many other works), and they are theoretically interesting by themselves.
They introduce possibilities, which are absent in continuum; for example, one
can consider discrete gauge groups as well as cotinuous ones. So far, the
main interest of people has been the study of lattice gauge theories
(specially pure gauge theories) on multidimensional lattices [2,6]. One can
not, however, consider the most general gauge theories on such lattices.

In the case of one-dimensional lattices, there is a dramatic change: there is
only one Wilson loop (in closed lattices). So one can consider the general
form of gauge-invariant interactions, including matter fields as well as
gauge fields. We will see that these theories are all exactly solvable in the
thermodynamic limit. An example is the Kazakov-Migdal induced gauge theory in
one dimension, studied by Caselle et. al. [7]. To be more specific, we
consider gauge-invariant Hamiltonians with compact gauge groups, and show
that a statistical system having such an intraction is exactly solvable in
the thermodynamic limit (section I). We further show that in this limit the
gauge degrees of freedom decouple from the matter degrees of freedom
(section I).Then we consider more special cases where the matter part of
interaction becomes a noninteracting theory, and observables become
uncorrelated (section II). In some cases one can also compute things for
finite lattices (section III). In section IV some examples are presented, and
in section V we consider double gauge field theories. These are all examples
of exactly solvable nonlocal interactions. There is no phase transition in
these theories, even at zero temperature. However, in a future paper, we will
study a generalized version of these theories which does have nontrivial
phase structure.

{\lbf O.I Gauge theory on lattices}

Consider a lattice consisting of a given set of sights $i$ and links $<ij>$.
Next consider two sets $V$ and $\widetilde V$, a function\hskip 3mm
$\widetilde{}$ : $ V\rightarrow\widetilde V$, and a mutipliaction from
$\widetilde V\times V$ to $\widetilde V V$. Defining the matter field $S$ on
sights, one can write the Hamiltonian for a nearest neighbour interaction as
$$H_{\circ}=-\sum_{<ij>}F(\widetilde{S_i}S_j)\eqno ({\rm O}.1)$$
where $F$ is a real-valued function.

Now, suppose that a group $G$ acts on the sets $V$ and $\widetilde V$
 through
$$\eqalignno{S&\rightarrow \hat g S& ({\rm O}.2)\cr
\widetilde{(\hat g S)}&=\widetilde S\hat g^{-1}& ({\rm O}.3)\cr}$$
where $\hat g$ is a representation of $g$. Introducing a group
element-valued field defined on links, one reaches a gauge-invariant
Hamiltonian
$$H_{\rm m}=-\sum_{<ij>}F(\widetilde{S_i}\hat U_{<ij>}S_j)\eqno ({\rm O}.4)$$
This Hamiltonian is invariant under local gauge transformation [4]
$$\eqalign{S_i&\rightarrow\hat g_iS_i\cr
 U_{<ij>}&\rightarrow g_iU_{<ij>}g_j^{-1}\cr}\eqno ({\rm O}.5)$$
To this (matter field) Hamiltonian, one can add another function, which is a
conjugation-invariant function (class function) of Wilson loops [4]
$$\eqalign{H&=-\sum_{<ij>}F(\widetilde{S_i}\hat U_{<ij>}S_j)
-E(W_{l_1},W_{l_2},\ldots )\cr &=:H_{\rm m}+H_G\cr}\eqno ({\rm O}.6)$$
where $W_l$s are Wilson loops.

Observables of this theory are of two kinds: gauge-invariant paths
$(\widetilde S_iW_{i\cdots j}S_j)$, $W_{i\cdots j}$ is a Wilson path
starting from $i$ ending in $j$, and class functions of Wilson loops.
Our task is to consider the partition function $Z$ and correlation
functions $<\Omega >$ of a statistical system, having an interaction
of the form ({\rm O}.6). These are defined as
$$Z:=\int\Big(\prod\limits_i dS_i\Big)\Big(\prod\limits_{<ij>}
dU_{<ij>}\Big)\exp\Big[-\beta H\big(\{ S_i\} ,\{U_{<ij>}\} \big)\Big]
\eqno ({\rm O}.7)$$
$$\Big<\Omega \big(\{ S_i\} ,\{ U_{<ij>}\} \big)\Big> :={1\over Z}\int
\Big(\prod\limits_i dS_i\Big)\Big(\prod\limits_{<ij>}
dU_{<ij>}\Big)\exp\Big[-\beta H\big(\{ S_i\} ,\{ U_{<ij>}\} \big)\Big]
\Omega \big(\{ S_i\} ,\{ U_{<ij>}\} \big)\eqno ({\rm O}.8)$$
In both cases the integration symbol is formal and may be integration or
summation, according to whether we have a discrete or continuous set. The
measure of group is the invariant measure, and the measure of matter field is
also invariant under the action of group.

{\lbf O.II One dimensional lattice gauge theory}

A one dimensional open lattice has no Wilson loop, and a closed one has only
one independent Wilson loop. So the Hamiltonian (O.6) is highly restricted
and we have (for closed lattices)
$$H=-\sum_{i=1}^N F(\widetilde S_i\hat U_{i+1/2}S_{i+1})-E\Big(\prod\limits_
{i=1}\limits^N U_{i+1/2}\Big)\eqno ({\rm O}.9)$$
where
$$\eqalignno{X_{N+k}&:=X_k& ({\rm O}.10)\cr\noalign{\hbox{and}}
U_{i+1/2}&:=U_{<i\; i+1>}& ({\rm O}.11)\cr}$$

In one dimension, the number of gauge degrees of freedom is exactly equal to
the number of gauge transformations. So it seems possible to eliminate the
gauge field by suitable gauge fixing. This is almost the case: for open
lattices this can be done and the Hamiltonian then reduces to $H_{\circ}$.
So, for open lattices gauging has no effect.

If the lattice is closed, however, one can not trivialize the Wilson loop by
gauge transformation. In fact, by a suitable gauge transformation,
$$\eqalignno{g_i&=\prod\limits_{k=1}\limits^iU_{k-1/2}\hskip 1in 1\leq i\leq
N& ({\rm O}.12)\cr U_{i+1/2}&\rightarrow\cases{1&$,i\neq
0$\cr\prod\limits_{k=1}\limits^NU_{k-1/2}&$,i=0$\cr}& ({\rm O}.13)\cr}$$
So the gauge field can be absorbed neither from the matter Hamiltonian nor
from the gauge part. However, we will show that, if the gauge group $G$ is
compact, in the thermodynamic limit $(N\rightarrow\infty )$ the Willson loop
has no effect on matter Hamiltonian and the partition function and
correlators can be factorized.

\noindent{\lbf I General results}

\noindent First of all, let us prove the factorizability of the partition
function and correlators. To do so, first consider a case where $G$ acts
transitively on $V$ (that is, $V$ consists of a single orbit of $G$).
Throughout this argument we assume that the group $G$ is compact.
We want to prove that, in the thermodynamic limit, the partial partition
function
$$Z_{\rm m}:=\int\Big(\prod\limits_i dS_i\Big)\exp\Big[\sum_i
f(\widetilde S_i\hat U_{i+1/2}S_{i+1})\Big]\eqno ({\rm I}.1)$$
is independent of $U_{i+1/2}$s ($f:=\beta F$). Defining the transfer
opertor $P(U)$ as
$$\big[\psi P(U)\big] (S):=\int dS'\psi (S')\exp\big[ f(\widetilde S'
\hat U S)\big]\eqno ({\rm I}.2)$$
it is obvious that
$$Z_{\rm m}={\rm tr}\Big[\prod\limits_i P(U_{i+1/2})\Big]
\eqno ({\rm I}.3)$$
Now consider the following lemma

{\it Lemma}: The eigenvector corresponding to the largest eigenvalue of
the transfer operator $P(U)$ is independent of $U$.

{\it proof}: From (I.2) we have
$$\big[\psi P(U)\big] (S)\leq{\rm max}\big\{\psi (S')\big\}\int dS'
\exp\big[ f(\widetilde S'\hat U S)\big]\eqno ({\rm I}.4)$$
and, as the integration measure is invariant under the action of $G$,
the integral in the right hand side of (I.4) does not depend on $U$ or
$S$. So, if $\psi$ obtains its maximum at $S_{\rm max}$, we have
$$\big[\psi P(U)\big] (S)\leq\mu\psi (S_{\rm max})\eqno ({\rm I}.5)$$
where $\mu$ is a constant:
$$\mu :=\int dS'\exp\big[ f(\widetilde S'\hat U S)\big]\eqno ({\rm I}.6)$$

Now suppose that $\psi$ is an eigenvector of $P(U)$ with eigenvalue
$\lambda$. That is
$$\big[\psi P(U)\big] (S)=\lambda\psi (S)\eqno ({\rm I}.7)$$
We then have
$$\lambda\psi (S_{\rm max})\leq\mu\psi (S_{\rm max})\eqno ({\rm I}.8)$$
One can always choose $\psi$ so that $\psi (S_{\rm max})$ is positive. Then,
from (I.8),
$$\lambda\leq\mu\eqno ({\rm I}.9)$$
and equality holds iff we have equality in (I.4), that is, iff
$$\psi (S)={\rm const.}\eqno ({\rm I}.10)$$

Therefor the eigenvector of $P(U)$ corresponding to its largest eigenvalue is
the constant function, which does not depend on $U$. The largest eigenvalue
does not depend on $U$ either.

Note that in the proof we have used the compactness of $G$, and hence $V$,
to guarantee the existence of a maximum for a continuous real-valued
function on $V$. We also notice that the largest eigenvalue $\mu$, defined
through (I.6), is indeed finite.

Using the above lemma, one can deduce that, in the thermodynamic limit,
$$Z_{\rm m}=\mu^N\eqno ({\rm I}.11)$$
So (setting $e:=\beta E$)
$$\eqalignno{Z&=\mu^N\int\Big(\prod\limits_i dU_{i+1/2}\Big)\exp\Big[
e\Big(\prod\limits_i U_{i+1/2}\Big)\Big]\cr&=\big[\mu{\rm vol}(G)\big]^N
{1\over{{\rm vol}(G)}}\int dU\exp\big[ e(U)\big]\cr
Z&=\big[\mu{\rm vol}(G)\big]^N\nu&({\rm I}.12)\cr}$$
where
$$\eqalignno{{\rm vol}(G)&=\int dU&({\rm I}.13)\cr
\nu:&={1\over{{\rm vol}(G)}}\int dU\exp\big[ e(U)\big]&({\rm I}.14)\cr}$$
{}From now on we will use a normalization for the group measure such that
the volume of the group becomes unity.

This argument can be readily generalized to cases where $V$ is not a
single orbit of $G$. In such a case we have
$$\eqalign{Z_{\rm m}&=\int\Big(\prod\limits_i dS_i\Big)\exp\Big[\sum_i
f(\widetilde S_i\hat U_{i+1/2}S_{i+1})\Big]\cr&=\int\Big(\prod\limits_i
dS_i\Big)\exp\Big[\sum_i f(\widetilde S_i\hat g_i^{-1}\hat U_{i+1/2}
\hat g_{i+1}S_{i+1})\Big]\cr&=\int\Big(\prod\limits_i dS_i\Big)\Big(
\prod\limits_i dg_i\Big)\exp\Big[\sum_i f(\widetilde S_i\hat g_i^{-1}
\hat U_{i+1/2}\hat g_{i+1}S_{i+1})\Big]\cr}\eqno ({\rm I}.15)$$
where we have used the invariance of $dS_i$ under the action of group. Now
define a partial transfer operator $P_G(U,S,S')$ by
$$(\psi P_G)(g):=\int dg'\psi (g')\exp\big[ f(\widetilde S'\hat g'^{-1}
\hat U \hat gS)\big]\eqno ({\rm I}.16)$$
We then have
$$Z_{\rm m}=\int\Big(\prod\limits_i dS_i\Big){\rm tr}\Big[\prod\limits_i
P_G(U_{i+1/2},S_{i+1},S_i)\Big]\eqno ({\rm I}.17)$$
Similarly, the eigenvector corresponding to the largest eigenvalue of $P_G$
is independent of $U$ and depends only on the orbits of $S$ and $S'$, which
we denote them by $\vert S\vert$ and $\vert S'\vert$. So, in the
thermodynamic limit we have
$$Z_{\rm m}=\int\Big(\prod\limits_i dS_i\Big)\Big[\prod\limits_i\mu \big(
\vert S_i\vert ,\vert S_{i+1}\vert \big)\Big] \eqno ({\rm I}.18)$$
where $\mu$ is the largest eigenvalue of $P_G$, and
$$Z=\nu\int\Big(\prod\limits_i dS_i\Big)\Big[\prod\limits_i\mu \big(
\vert S_i\vert ,\vert S_{i+1}\vert \big)\Big]\eqno ({\rm I}.19)$$

What about the correlation functions? This argument still works if we are
considering correlators of observables confined to a finite region. By a
suitable gauge fixing, one can eliminate the gauge fields in that region.
Then the above arguments are valid and, in the thermodynamic limit, the
integral
$$\int\Big(\prod\limits_i dS_i\Big)\Omega_{\rm fin.}\big(\{ S_i\} ,
\{ U_{i+1/2}\}\big)\exp (-\beta H_{\rm m})$$
is independent of $U_{i+1/2}$s. So we have
$$\eqalign{\big<\Omega_{\rm fin.}\big>&={1\over{\nu Z_{\rm m}}}
\nu\int\Big(\prod\limits_i dS_i\Big)\Omega_{\rm fin.}\exp (-\beta H_{\rm m})
\cr &={1\over Z_{\rm m}}\int\Big(\prod\limits_i dS_i\Big)\Omega_{\rm fin.}
\exp (-\beta H_{\rm m})\cr}\eqno ({\rm I}.20)$$
and, as the right hand sides are independent of $U_{i+1/2}$s,
$$\eqalign{\big<\Omega_{\rm fin.}\big>&=\big<\Omega_{\rm fin.}
\big>_{H_{\rm m}}\cr &=\big<\Omega^{\circ}_{\rm fin.}\big>_{H_{\circ}}\cr}
\eqno ({\rm I}.21)$$
where by subscripts $H_{\rm m}$ and $H_{\circ}$ we mean averaging with
Boltzman weights corresponding to $H_{\rm m}$ and $H_{\circ}$, respectively,
and
$$\Omega^{\circ}_{\rm fin.}\big(\{ S_i\} ,\{ U_{i+1/2}\}\big) :=
\Omega_{\rm fin.}\big(\{ S_i\} ,\{ U_{i+1/2}=1\}\big)\eqno ({\rm I}.22)$$

Now suppose that $\Omega$ is the product of $\Omega_G$ and
$\Omega_{\rm fin.}$, where $\Omega_G$ is a function of the Wilson loop. It
is easily seen that,
$$\eqalign{\big<\Omega\big>&=\big<\Omega_G\big>\big<\Omega_{\rm fin.}\big>
\cr &=\big<\Omega_G\big>_{H_G}\big<\Omega^{\circ}_{\rm fin.}\big>_{H_{\circ}}
\cr}\eqno ({\rm I}.23)$$

So we have proved that\hfil\break
{\it A one dimensional gauge theory with nearest neighbour interaction
between matter fields on a closed lattice and with a compact gauge group
is (as long as we are cosidering observables which are either local or
functions of the Wilson loop), in the thermodynamic limit, effectively
decomposed to two noninteracting parts: A matter part, the Hamiltonian
of which is $H_{\rm m}$ (or equivalently $H_{\circ}$), and a gauge part,
which is a one particle system.}\vfil\break

\noindent{\lbf II Matter field spaces consisting of a single orbit of $G$}

\noindent In this case, one can completely eliminate  the matter field by
suitable gauge fixing (even for finite lattices). In fact, the partial
partition function
$$Z_{\rm p}:=\int\Big(\prod\limits_i dU_{i+1/2}\Big)\exp\Big[\sum_i
f(\widetilde S_i\hat U_{i+1/2}S_{i+1})+e\Big(\prod\limits_i U_{i+1/2}\Big)
\Big]\eqno ({\rm II}.1)$$
is independent of $S_i$s. To see this, suppose that we change $S_k$ to
$$S'_k=\hat gS_k\eqno ({\rm II}.2)$$
leaving other $S_i$s unchanged. We then have
$$Z_{\rm p}(S'_k)=\int\Big(\prod\limits_i dU'_{i+1/2}\Big)\exp\Big[\sum_i
f(\widetilde S_i\hat U'_{i+1/2}S_{i+1})+e\Big(\prod\limits_i U'_{i+1/2}\Big)
\Big]\eqno ({\rm II}.3)$$
where we have made the change of variable
$$U'_{k-1/2}=U_{k-1/2}\;g\hskip 2cm ,\hskip 2cm U'_{k+1/2}=g^{-1}U_{k+1/2}
\eqno ({\rm II}.4)$$
So
$$Z_{\rm p}\big(\{ S'_i\}\big)=Z_{\rm p}\big(\{ S_i\}\big)
\eqno ({\rm II}.5)$$
It is easily seen that this argument can also be generalized to
correlators. So, we can eliminate the matter field and use the gauge-fixed
Hamiltonian
$$H_{\rm gf}:=-\sum_i F(\widetilde C\hat U_{i+1/2}C)-E\Big(\prod\limits_i
U_{i+1/2}\Big)\eqno ({\rm II}.6)$$
where $C$ is a constant member of $V$. Using the result of section I,
the theory is decomposed to two noninteracting parts (in the thermodynamic
limit):
$$(H_{\rm m})_{\rm gf}:=-\sum_i F(\widetilde C\hat U_{i+1/2}C)
\eqno ({\rm II}.7)$$
and $H_G$. But now, there is no interaction in the matter part. So we
conclude that, as long as we are considering local observables, the
theory is free and essentially a one-particle theory. That is, observables
at different points are uncorrelated, and of course distance-independent:
$$\big<\Omega_1\Omega_2\big> =\big<\Omega_1\big>\big<\Omega_2\big>
\eqno ({\rm II}.8)$$
if $\Omega_1$ and $\Omega_2$ depend on no common $U_i$. We also have
$$\big<\Omega\big> =\big<\Omega\big>_{(H_{\scriptstyle{\rm m}})
_{{\scriptstyle{\rm gf,r(\Omega )}}}}\eqno ({\rm II}.9)$$
where $(H_m)_{\rm gf,r(\Omega )}$ is the sum of those terms of
$(H_m)_{\rm gf}$ in them $U_i$s contributing in $\Omega$ enter. In the above
arguments, $\Omega$ is a function of gauge-invariant paths with the
substitution $S_i\rightarrow C$.

The total partition function takes the form
$$Z=\Big\{\Big(\int dS\Big)\int dU\exp\big[ f(\widetilde C\hat UC)\big]
\Big\}^N \int dU\exp\big[ e(U)\big]\eqno ({\rm II}.10)$$
comparing this with (I.12), we find
$${\rm vol}(V)\int dU\exp\big[ f(\widetilde C\hat UC)\big] =
{\rm vol}(G)\int dS'\exp\big[ f(\widetilde S'\hat US)\big]
\eqno ({\rm II}.11)$$

So any correlation function can be obtained by a (finite) number of
integrations. As an example consider pure-gauge correlators, that is
class functions of the Wilson loop. These functions are linear combinations
of group characters (see the appendix); so one only needs to consider
correlators of the form
$$\Big< X_{\mu}\Big(\prod\limits_i U_{i+1/2}\Big)\Big>={{\int dU\; X_{\mu}(U)
\exp\big[ e(U)\big]}\over{\int dU\exp\big[ e(U)\big]}}\eqno ({\rm II}.12)$$
where $X_{\mu}$ is the character of group in the (unitary)
representation $\mu$. $\exp\big[ e(U)\big]$ itself is also a class function,
so one can expand it as
$$\eqalign{\exp\big[ e(U)\big]&=\exp\big[\beta E(U)\big]\cr
&=\sum_{\lambda}I_{\lambda}^{(G,E)}(\beta )X_{\lambda}(U)\cr}
\eqno ({\rm II}.13)$$
where the summation runs over irreducible unitary representations of $G$,
and
$$I_{\lambda}^{(G,E)}(\beta )=\int dU\; X_{\lambda}(U^{-1})\exp
\big[\beta E(U)\big]\eqno ({\rm II}.14)$$
(See the appendix). From these we obtain
$$\Big< X_{\mu}\Big(\prod\limits_i U_{i+1/2}\Big)\Big> ={{I_{\bar\mu}^{(G,E)}
(\beta )}\over{I_0^{(G,E)}(\beta )}}\eqno ({\rm II}.15)$$
where $\bar\mu$ is the complex conjugate representation of $\mu$, and
0 is the trivial representation. There is a special case where one
can go further and calculate correlators for finite lattices. We will
consider this case in the following section.

\noindent{\lbf III Conjugation-invariant matter Hamiltonians and observables}

\noindent As a special case of section II, consider a gauge-fixed matter
Hamiltonian having the following property
$$F(\widetilde C\hat UC)=F(\widetilde C\hat g^{-1}\hat U\hat gC)
\eqno ({\rm III}.1)$$
If we restrict ourselves to observables which have the same property,
that is invariance under local conjugation, we can compute correlators for
finite lattices. First we compute the partition function:
$$Z=\int\Big(\prod\limits_i dU_{i+1/2}\Big)\exp\Big[\sum_i f_C(U_{i+1/2})+
e\Big(\prod\limits_i U_{i+1/2}\Big)\Big]$$
where
$$f_C(U):=f(\widetilde CUC)\eqno ({\rm III}.2)$$
using
$$\exp\big[\beta F_C(U)\big]=\sum_{\lambda}I_{\lambda}^{(G,F_C)}(\beta )
X_{\lambda}(U)\eqno ({\rm III}.3)$$
we have
$$Z=\sum_{\{\lambda_{i+1/2}\} ,\mu}\Big\{\int\Big(\prod\limits_i dU_{i+1/2}
\Big)\Big[\prod\limits_i X_{\lambda_{i+1/2}}(U_{i+1/2})\Big]X_{\bar\mu}
\Big(\prod\limits_i U_{i+1/2}\Big)\Big\} I_{\lambda_{i+1/2}}^{(G,F_C)}
(\beta )I_{\bar\mu}^{(G,E)}(\beta )\eqno ({\rm III}.4)$$
and (using the appendix)
$$\int\Big(\prod\limits_i dU_{i+1/2}\Big)\Big[\prod\limits_i
X_{\lambda_{i+1/2}}(U_{i+1/2})\Big] X_{\bar\mu}\Big(\prod\limits_i
U_{i+1/2}\Big) =\Big(\prod\limits_i \delta_{\lambda_{i+1/2},\mu}\Big)
\big[ d(\mu )\big]^{1-N}\eqno ({\rm III}.5)$$
where $d(\mu )$ is the dimension of the representation $\mu$. So
$$Z=\sum_{\mu }\Big[{{I_{\mu}^{(G,F_C)}(\beta )}\over{d(\mu )}}\Big]^N
I_{\bar\mu}^{(G,E)}(\beta )d(\mu )\eqno ({\rm III}.6)$$

Now, consider correlators of the form
$$\Big<\prod\limits_i X_{\sigma_{i+1/2}}(U_{i+1/2})\Big>=:
\big<\{\sigma_{i+1/2}\}\big>\eqno ({\rm III}.7)$$
Every (conjugation-invariant) correlator can be expanded in terms of these.
We have
$$\eqalign{\big<\{\sigma_{i+1/2}\}\big>=&{1\over Z}
\sum_{\{\lambda_{i+1/2}\} ,\mu}\Big\{\int\Big(\prod\limits_i dU_{i+1/2}\Big)
\Big[\prod\limits_i X_{\sigma_{i+1/2}}(U_{i+1/2})\; X_{\lambda_{i+1/2}}
(U_{i+1/2})\Big] X_{\bar\mu}\Big(\prod\limits_i U_{i+1/2}\Big)\Big\}\cr
&I_{\lambda_{i+1/2}}^{(G,F_C)}(\beta )I_{\bar\mu}^{(G,E)}(\beta )\cr}
\eqno ({\rm III}.8)$$
Defining nonnegative integers $\{{\rho ,\tau ;\phi}\}$ through
$$[\rho ]{\scriptscriptstyle\bigotimes}[\tau ]={\scriptscriptstyle
{\bigoplus\limits_{\displaystyle\phi}}} \{\rho ,\tau ;\phi\} [\phi ]
\eqno ({\rm III}.9)$$
where $[\rho ]$, $[\tau ]$, and $[\phi ]$ are irreducible unitary
representations of $G$, we conclude that
$$\big<\{\sigma_{i+1/2}\}\big>={1\over Z}\sum_{\mu}\bigg\{\prod\limits_i
\Big[{{\sum_{\lambda_{i+1/2}}\{\sigma_{i+1/2},\lambda_{i+1/2};\mu\}
I_{\lambda_{i+1/2}}^{(G,F_C)}(\beta )}\over{d(\mu )}}\Big]\bigg\}
I_{\bar\mu}^{(G,E)}(\beta )d(\mu )\eqno ({\rm III}.10)$$

We see, as one expects from the very beginning, that these correlators
are distance-independent. That is, they depend only on the set of
$\sigma_{i+1/2}$s, not on their ordering. One can also calculate the
correlator for the characters of the Wilson loop. The calculation is
similar to above, and one obtains
$$\Big< X_{\sigma}\Big(\prod\limits_i U_{i+1/2}\Big)\Big>={1\over Z}
\sum_{\mu ,\lambda}\{\sigma ,\mu ;\bar\lambda\} I_{\mu}^{(G,E)}(\beta )
\; d(\lambda )\Big[{{I_{\lambda}^{(G,F_C)}(\beta )}\over{d(\lambda )}}\Big]^N
\eqno ({\rm III}.11)$$

One can also easily go to the thermodynamic limit. Using
$$\big\vert X_{\lambda}(U)\big\vert\leq d(\lambda )\hskip 1.5cm
{\hbox{if $\lambda\neq 0$}}\eqno ({\rm III}.12)$$
and
$$X_0(U)=1\eqno ({\rm III}.13)$$
it is seen that
$$\Big\vert{{I_{\lambda}^{(G,F_C)}(\beta )}\over{d(\lambda )}}\Big\vert
<I_0^{(G,F_C)}(\beta )\hskip 1.5cm {\hbox{if $\lambda\neq 0$}}
\eqno ({\rm III}.14)$$
So, in the limit $N\rightarrow\infty$,
$$Z=\big[ I_0^{(G,F_C)}(\beta )\big]^NI_0^{(G,E)}(\beta )
\eqno ({\rm III}.15)$$
$$\eqalign{\big<\{\sigma_{i+1/2}\}\big>&={1\over Z}\Big\{\prod\limits_i
\Big[\sum_{\lambda_{i+1/2}}\{\sigma_{i+1/2},\lambda_{i+1/2};0\}
I_{\lambda_{i+1/2}}^{(G,F_C)}(\beta )\Big]\Big\}
I_0^{(G,E)}(\beta )\cr &={1\over Z}\Big[\prod\limits_i
I_{\bar\sigma_{i+1/2}}^{(G,F_C)}(\beta )\Big] I_0^{(G,E)}(\beta )
\cr}$$
or
$$\eqalign{\big<\{\sigma_{i+1/2}\}\big>&=\prod\limits_i
{{I_{\bar\sigma_{i+1/2}}^{(G,F_C)}(\beta )}\over{I_0^{(G,F_C)}(\beta )}}
\cr &=\prod\limits_i\big< X_{\sigma_{i+1/2}}(U_{i+1/2})\big>\cr}
\eqno ({\rm III}.16)$$
Note that we have assumed that the observable is local, i.e. only a finite
number of $\sigma_{i+1/2}$s are nonzero.

Finally
$$\eqalign{\Big< X_{\sigma}\Big(\prod\limits_i U_{i+1/2}\Big)\Big>
&={1\over Z}\sum_{\mu}\{\sigma ,\mu ;0\}I_{\mu}^{(G,E)}(\beta )
\big[I_0^{(G,F_C)}\big]^N\cr
&={{I_{\bar\sigma}^{(G,E)}(\beta )}\over{I_0^{(G,E)}(\beta )}}
\cr}\eqno ({\rm III}.17)$$

One can see that in this limit observables at different points are
uncorrelated, and the theory is factorized into a matter part and a pure
gauge part, just as from the previous theorems is deduced. Also note that
if $G$ is Abelian, the condition of conjugation invariance is automatically
satisfied. So, all of the results obtained in this section are valid.

\noindent{\lbf IV Examples}

{\lbf IV.I Gauge-invariant Ising model}

As the simplest case, consider
$$H_{\circ}=-J\sum_i S_i S_{i+1}\eqno ({\rm IV}.1)$$
where each $S_i$ takes the values $\pm 1$. $H_{\circ}$ has global gauge
symmetry under the action of gauge group $Z_2$. From this, one can construct
the gauge-invariant Hamiltonian
$$H=-J\sum_i S_i U_{i+1/2}S_{i+1} -K\prod\limits_i U_{i+1/2}
\eqno ({\rm IV}.2)$$
where $U_{i+1/2}$s also take the values $\pm 1$.

This group has only two representations: the defining repersentation, and
the trivial one. So, from (III.6), we have
$$Z=\sum_{\lambda =0}^1\big[I_{\lambda}^{(Z_2)}(\beta J)\big]^N
I_{\lambda}^{(Z_2)}(\beta K)$$
where
$$I_0^{(Z_2)}(x):={1\over 2}\sum_{U=\pm 1}\exp (xU)=\cosh x
\eqno ({\rm IV}.3)$$
and
$$I_1^{(Z_2)}(x):={1\over 2}\sum_{U=\pm 1}U\exp (xU)=\sinh x
\eqno ({\rm IV}.4)$$
So,
$$Z=\cosh^N(\beta J)\cosh (\beta K)+\sinh^N(\beta J)\sinh (\beta K)
\eqno ({\rm IV}.5)$$
Using
$$\{\rho ,\tau ;\phi\} =\delta^{(2)}_{\phi ,\rho +\tau}\eqno ({\rm IV}.6)$$
where
$$\delta^{(n)}_{m,l}:=\cases{1,&$m\equiv l \pmod n$\cr 0,&
 otherwise\cr}\eqno ({\rm IV}.7)$$
$$\big<\{\sigma_{i+1/2}\}\big> ={1\over Z}\Big[ I_0^{(Z_2)}
(\beta K)\prod\limits_i I_{\sigma_{i+1/2}}^{(Z_2)}(\beta J)+I_1^{(Z_2)}
(\beta K)\prod\limits_i I_{1-\sigma_{i+1/2}}^{(Z_2)}(\beta J)\Big]$$
or
$$\big<\{\sigma_{i+1/2}\}\big> ={{\cosh (\beta K)\cosh^N(\beta J)
\tanh^s(\beta J)+\sinh (\beta K)\sinh^N(\beta J)\tanh^{-s}(\beta J)}
\over{\cosh (\beta K)\cosh^N(\beta J)+\sinh (\beta K)\sinh^N(\beta J)}}
\eqno ({\rm IV}.8)$$
where
$$s:=\sum_i\sigma_{i+1/2}\eqno ({\rm IV}.9)$$
And finally
$$\Big<\prod\limits_i U_{i+1/2}\Big>={{\cosh (\beta K)\sinh^N(\beta J)+
\sinh (\beta K)\cosh^N(\beta J)}\over{\cosh (\beta K)\cosh^N(\beta J)+
\sinh (\beta K)\sinh^N(\beta J)}}\eqno ({\rm IV}.10)$$
For $N\rightarrow\infty$,
$$\eqalignno{Z&=\cosh (\beta K)\cosh^N(\beta J)&({\rm IV}.11)\cr
\big<\{\sigma_{i+1/2}\}\big>&=\tanh^s(\beta J)&({\rm IV}.12)\cr
\Big<\prod\limits_i U_{i+1/2}\Big>&=\tanh (\beta K)&({\rm IV}.13)\cr}$$

{\lbf IV.II Gauge-invariant Potts model}

$$H_{\circ}=-2J\sum_i\big(\delta_{S_i,S_{i+1}}-{1\over 2}\big)
\eqno ({\rm IV}.14)$$
where each $S_i$ is a nonnegative integer less than $n$. This Hamiltonian
has global $Z_n$ invariance. The gauge-invariant Hamiltonian is
$$H=-2J\sum_i\big[\delta^{(n)}_{S_{i+1}+U_{i+1/2}-S_i,0}-{1\over 2}\big]
-E\Big(\sum_i U_{i+1/2}\Big)\eqno ({\rm IV}.15)$$
where $E$ is a periodic function on integers with period $n$. $Z_n$ has
$n$ representations, the characters of them are
$$X_{\lambda}(m):=\exp\Big({{i2\pi\lambda m}\over n}\Big)
\eqno ({\rm IV}.16)$$
And we have
$$\eqalignno{I_{\lambda}^{(Z_n,\delta )}(x)&:={1\over n}\sum_{m=0}^{n-1}
\exp\Big( -{{i2\pi\lambda m}\over n}\Big)\exp\big( x\delta_{m,0}\big)
\hskip 1cm\Rightarrow\cr I_{\lambda}^{(Z_n,\delta )}(x)&={{\exp (x)-1}
\over n}+\delta_{\lambda ,0}&({\rm IV}.17)\cr}$$
So
$$Z=\bigg\{ I_0^{(Z_n,E)}(\beta )\Big[ 1+{{\exp (2\beta J)-1}\over n}
\Big]^N+\Big[\sum_{\lambda =1}^{n-1}I_{n-\lambda}^{(Z_n,E)}(\beta )\Big]\Big[
{{\exp (2\beta J)-1}\over n}\Big]^N\bigg\}\exp (-N\beta J)$$
or
$$Z=\Big({2\over n}\Big)^N\bigg\{ I_0^{(Z_n,E)}(\beta )\Big[
{n\over 2}\exp (-\beta J)+\sinh (\beta J)\Big]^N+\sinh^N(\beta J)
\sum_{\lambda =1}^{n-1}I_{n-\lambda}^{(Z_n,E)}(\beta )\bigg\}
\eqno ({\rm IV}.18)$$
where we have
$$I_{\lambda}^{(Z_n,E)}(x):={1\over n}\sum_{m=0}^{n-1}\exp
\Big( -{{i2\pi\lambda m}\over n}\Big)\exp\big[ xE(m)\big]
\eqno ({\rm IV}.19)$$
$$\big<\{\sigma_{i+1/2}\}\big> ={{(2/n)^N}\over Z}\sum_{\lambda =0}
^{n-1}\bigg\{I_{n-\lambda}^{(Z_n,E)}(\beta )\prod\limits_i\Big[{n\over 2}
\exp (-\beta J)\delta_{\lambda ,\sigma_{i+1/2}}+\sinh (\beta J)\Big]\bigg\}
\eqno ({\rm IV}.20)$$
$$\Big< X_{\sigma}\Big(\prod\limits_i U_{i+1/2}\Big)\Big>={{(2/n)^N}\over Z}
\bigg\{ I_{n-\sigma}^{(Z_n,E)}(\beta )\Big[{n\over 2}\exp (-\beta J)
+\sinh (\beta J)\Big]^N+\sinh^N(\beta J)\sum_{\lambda =1}^{n-1}
I_{n-\lambda -\sigma}^{(Z_n,E)}(\beta )\bigg\}\eqno ({\rm IV}.21)$$
One can easily verify that these relations reduce to those of Ising model,
if $n=2$ and $E(x)=K\cos (x)$.

In the thermodynamic limit one obtains
$$\eqalignno{Z&=\Big({2\over n}\Big)^NI_0^{(Z_n,E)}(\beta )
\Big[{n\over 2}\exp (-\beta J)+\sinh (\beta J)\Big]^N&({\rm IV}.22)\cr
\big<\{\sigma_{i+1/2}\}\big>&=\Big[{{\sinh (\beta J)}\over{(n/2)
\exp (-\beta J)+\sinh (\beta J)}}\Big]^s&({\rm IV}.23)\cr
\Big< X_{\sigma}\Big(\prod\limits_i U_{i+1/2}\Big)\Big>&={{I_{n-\sigma}
^{(Z_n,E)}(\beta )}\over{I_0^{(Z_n,E)}(\beta )}}&({\rm IV}.24)\cr}$$
where we have
$$s:=\sum_i\big( 1-\delta_{\sigma_{i+1/2},0}\big)\eqno ({\rm IV}.25)$$

{\lbf IV.III Gauge-invariant classical planar spin model}

Taking
$$H_{\circ}=-J\sum_i{\rm Im}(S_i^* S_{i+1})\eqno ({\rm IV}.26)$$
where $S_i$s are phases, we see that the gauge group is $U(1)$. The
gauge-invariant Hamiltonian is
$$H=-J\sum_i{\rm Im}(S_i^* U_{i+1/2}S_{i+1})-E\Big(\prod_i U_{i+1/2}\Big)
\eqno ({\rm IV}.27)$$
where $E$ is a function from $U(1)$ to $R$. We parametrize the gauge group
by $\xi$. There are (countably) infinite representations labeled by integers:
$$\hat U_{\lambda}(\xi )=\exp (i\lambda\xi )\eqno ({\rm IV}.28)$$
And we have
$$\eqalign{I_{\lambda}^{\left[ U\left( 1\right) ,J\;{\rm Im}\right]}(\beta )
&:={1\over{2\pi}}\int_0^{2\pi}d\xi\exp (\beta J\sin\xi -i\lambda\xi )\cr
&=i^{-\lambda}I_{\lambda}(\beta J)\cr}\eqno ({\rm IV}.29)$$
where $I_{\lambda}(x)$ is the modified Bessel function. We also have
$$\{\rho ,\tau ;\phi\} =\delta_{\phi ,\rho +\tau}\eqno ({\rm IV}.30)$$
So
$$\eqalignno{Z&=\sum_{\lambda =-\infty}^{+\infty}\big[ i^{-\lambda}
I_{\lambda}(\beta J)\big]^NI_{-\lambda}^{\left[ U\left( 1\right) ,E\right]}
(\beta )&({\rm IV}.31)\cr \big<\{\sigma_{i+1/2}\}\big>&={1\over Z}
\sum_{\lambda =-\infty}^{+\infty}\Big\{I_{-\lambda}^{\left[ U\left( 1\right)
,E\right]}(\beta )\prod\limits_i\big[ i^{-\lambda +\sigma_{i+1/2}}
I_{\lambda -\sigma_{i+1/2}}(\beta J)\big]\Big\}&({\rm IV}.32)\cr
\Big<\exp\Big[ i\sigma\Big(\sum_i\xi_{i+1/2}\Big)\Big]\Big>&={1\over Z}
\sum_{\lambda =-\infty}^{+\infty}\big[ i^{-\lambda}I_{\lambda}(\beta J)\big]
^NI_{-\lambda -\sigma}^{\left[ U\left( 1\right) ,E\right]}(\beta )
&({\rm IV}.33)\cr}$$
And, in the thermodynamic limit
$$\eqalignno{Z&=\big[ I_0(\beta J)\big]^NI_0^{\left[ U\left( 1\right)
,E\right]}(\beta )&({\rm IV}.34)\cr \big<\{\sigma_{i+1/2}\}\big>
&=\prod\limits_i\Big[{{i^{\sigma_{i+1/2}}I_{\sigma_{i+1/2}}(\beta J)}\over
{I_0(\beta J)}}\Big]&({\rm IV}.35)\cr \Big<\exp\Big[ i\sigma\Big(
\sum_i\xi_{i+1/2}\Big)\Big]\Big>&={{I_{-\sigma}^{\left[ U\left( 1\right)
,E\right]}(\beta )}\over{I_0^{\left[ U\left( 1\right) ,E\right]}(\beta )}}
&({\rm IV}.36)\cr}$$

{\lbf IV.IV $Z_2$ gauge theory on multiple orbit matter field space}

As our last example consider a generalized form of gauge-invariant Ising
model, where $S_i$s can take a set of values $\{ a_m\}$, which are symmetric
with respect to zero and have absolute values less than or equal to unity.
The gauge-invariant Hamiltonian is of the form (IV.2). Using the results of
section I, we have
$$\eqalign{\mu\big(\vert S_i\vert ,\vert S_{i+1}\vert\big) &:={1\over 2}
\Big\{\exp\big[\beta J(S_i S_{i+1})\big] +\exp\big[ -\beta J(S_i S_{i+1})
\big]\Big\}\cr &=\cosh (\beta JS_i S_{i+1})\cr}\eqno ({\rm IV}.37)$$
$$Z_{\rm m}=\sum_{\{ S_i\}}\cosh (\beta JS_i S_{i+1})\eqno ({\rm IV}.38)$$
So $Z_{\rm m}$ is (in the thermodynamic limit) the $N$th power of the
largest eigenvalue of the matrix $M$ with the entries
$$M_{pq}:=\cosh (\beta Ja_pa_q)\eqno ({\rm IV}.39)$$
For example,
$$\{ a_m\} =\{ 0,\pm 1\}\hskip 1cm\Rightarrow$$
$$\lambda_{\rm max}={{1+2\cosh (\beta J)+\sqrt{4\cosh^2(\beta J)-4\cosh
(\beta J)+9}}\over 2}\eqno ({\rm IV}.40)$$
$$\{ a_m\} =\{\pm 1/3,\pm 1\}\hskip 1cm\Rightarrow$$
$$\lambda_{\rm max}=\cosh (\beta J)+\cosh (\beta J/9)+\sqrt{\big[\cosh
(\beta J)-\cosh (\beta J/9)\big]^2+4\cosh^2(\beta J/3)}\eqno ({\rm IV}.41)$$
And in either case,
$$Z=\cosh (\beta K)(\lambda_{\rm max})^N\eqno ({\rm IV}.42)$$

Finally, if the set of possible values of $S_i$s is the interval $[-1,1]$,
we must solve the eigenvalue problem
$$\lambda\Lambda (x)=\int_{-1}^1\cosh (\beta Jxy)\Lambda (y)dy
\eqno ({\rm IV}.43)$$
for its largest eigenvalue, and insert it in (IV.42).

\noindent{\lbf V Double gauge field theories}

\noindent There is a special case where the theory has more invariance
properties. Consider a Hamiltonian
$$H_{\circ}=-\sum_{<ij>}F(S_i^{-1}S_j)\eqno ({\rm V}.1)$$
where $S_i$s themselves belong to a non-Abelian group $G$, and $F$ is a
real-valued class function on $G$. $H_{\circ}$ has two independent global
symmetries:
$$S_i\rightarrow g_L\; S_i,\hskip 1cm S_i\rightarrow S_i\;g_R
\eqno ({\rm V}.2)$$
where $g_L$ and $g_R$ are group elements. Introducing two gauge fields $L$
and $R$, one can make both of these invariances local. The general form of
gauge transformation is
$$\eqalign{S_i&\rightarrow (g_L)_iS_i(g_R)_i\cr
L_{<ij>}&\rightarrow (g_L)_iL_{<ij>}(g_L)_j^{-1}\cr
R_{<ij>}&\rightarrow (g_R)_i^{-1}R_{<ij>}(g_R)_j\cr}\eqno ({\rm V}.3)$$
It is easy to see that
$$H_{\rm m}=-\sum_{<ij>}F(S_i^{-1}L_{<ij>}S_jR_{<ji>})\eqno ({\rm V}.4)$$
is gauge-invariant. We can also add class functions of Wilson loops to the
above expression. In one dimension, one comes then to the Hamiltonian
$$H=-\sum_i F(S_i^{-1}L_{i+1/2}S_{i+1}R_{i+1/2})-E\Big( \prod
\limits_i^{\rightarrow}L_{i+1/2},\prod\limits_i^{\leftarrow}R_{i+1/2}\Big)
\eqno ({\rm V}.5)$$
where
$$\eqalignno{L_{i+1/2}&:=L_{<i\; i+1>}&({\rm V}.6)\cr
R_{i+1/2}&:=R_{<i+1\; i>}&({\rm V}.7)\cr
\prod\limits_i^{\rightarrow}X_i&:=X_1X_2\cdots&({\rm V}.8)\cr
\prod\limits_i^{\leftarrow}Y_i&:=\cdots Y_2Y_1&({\rm V}.9)\cr}$$
By gauge-fixing, one can eliminate $S_i$s and come to
$$H_{\rm gf}=-\sum_iF(L_{i+1/2}R_{i+1/2})-E\Big( \prod\limits_i
^{\rightarrow}L_{i+1/2},\prod\limits_i^{\leftarrow}R_{i+1/2}\Big)
\eqno ({\rm V}.10)$$
{}From which (using the appendix) we find
$$Z=\sum_{\mu}\Big[{{I_{\mu}^{(G,F)}(\beta )}\over{d(\mu )}}\Big]^N
I_{\bar\mu ,\bar\mu}^{(G,E)}(\beta )\eqno ({\rm V}.11)$$
where
$$\exp\big[\beta E(L,R)\big] =:\sum_{\lambda_L,\lambda_R}I_
{\lambda_L,\lambda_R}^{(G,E)}(\beta )X_{\lambda_L}(L)X_{\lambda_R}(R)
\eqno ({\rm V}.12)$$

Observables of this theory are
$$\prod\limits_i\big[X_{\sigma_{i+1/2}}(S_i^{-1}L_{i+1/2}S_{i+1}R_{i+1/2})
\big] ,\;\; {\rm and}\;\; X_{\sigma_L}\Big(\prod\limits_i^{\rightarrow}
L_{i+1/2}\Big) X_{\sigma_R}\Big(\prod\limits_i^{\leftarrow}R_{i+1/2}\Big).$$
In a manner similar to the previous cases one finds
$$\big<\{\sigma_{i+1/2}\}\big>={1\over Z}\sum_{\mu}\bigg\{
I_{\bar\mu ,\bar\mu}^{(G,E)}(\beta )\prod\limits_i\Big[{1\over{d(\mu )}}
\sum_{\lambda_{i+1/2}}\{\sigma_{i+1/2} ,\lambda_{i+1/2} ;\mu\}
I_{\lambda_{i+1/2}}^{(G,F)}(\beta )\Big]\bigg\}\eqno ({\rm V}.13)$$
$$\Big<X_{\sigma_L}\Big(\prod\limits_i^{\rightarrow}L_{i+1/2}\Big)
X_{\sigma_R}\Big(\prod\limits_i^{\leftarrow}R_{i+1/2}\Big)\Big>={1\over Z}
\sum_{\lambda ,\mu_L,\mu_R}\{\sigma_L,\mu_L;\bar\lambda\}
\{\sigma_R,\mu_R;\bar\lambda\}I_{\mu_L,\mu_R}^{(G,E)}(\beta )\Big[
{{I_{\lambda}^{(G,F)}(\beta )}\over{d(\lambda )}}\Big]^N\eqno ({\rm V}.14)$$
We see that the matter field correlations are distance-independent. In the
thermodynamic limit,
$$\eqalignno{Z&=\big[I_0^{(G,F)}(\beta )\big]^NI_{0,0}^{(G,E)}(\beta )
&({\rm V}.15)\cr \big<\{\sigma_{i+1/2}\}\big>&=\prod\limits_i\Big[
{{I_{\bar\sigma_{i+1/2}}{(G,F)}(\beta )}\over{I_0{(G,F)}(\beta )}}\Big]
&({\rm V}.16)\cr\Big< X_{\sigma_L}\Big(\prod\limits_i^{\rightarrow}
L_{i+1/2}\Big) X_{\sigma_R}\Big(\prod\limits_i^{\leftarrow}R_{i+1/2}\Big)
\Big>&={{I_{\bar\sigma_L,\bar\sigma_R}^{(G,E)}(\beta )}\over{I_{0,0}^{(G,F)}
(\beta )}}&({\rm V}.17)\cr}$$

In fact, in the thermodynamic limit we are faced with two decoupled
interactions: a one particle interaction, and a two particle one.

\noindent{\lbf VI Conclusion}

\noindent we have shown that a one dimensional gauge theory for compact gauge
groups is exactly solvable in the thermodynamic limit. In fact, in this limit
the gauge degrees of freedom completely decouple from the matter degrees of
freedom. The gauge part then reduces to a one particle theory. If the gauge
group acts transitively on matter field spaces, then the matter part also
reduces to a noninteracting system. In such a case, all of the observables
will be uncorrelated. We saw that there are certain cases, where the
correlators become distance-independent even for finite lattices. So far,
there has been no phase transition in these theories. However, in a future
paper, we will introduce a generalization of these theories which does
exhibit first order phase transition.

\noindent{\lbf VII Appendix}

\noindent Here we introduce a couple of group identities used in the text.
Take a compact group $G$, and let $X_{\lambda}(g)$ be the character (trace)
of the element $g$ in the (irreducible unitary) representation $\lambda$. We
have
$$\eqalign{X_{\bar\lambda}(g)&=\big[ X_{\lambda}(g)\big]^*\cr &=
X_{\lambda}(g^{-1})\cr}\eqno ({\rm VII}.1)$$
where $\bar\lambda$ is the complex conjugate representation of $\lambda$.
$$\int dg\;X_{\lambda}(hg)X_{\mu}(h'g^{-1})={{X_{\lambda}(hh')}\over
{d(\lambda )}}\delta_{\lambda ,\mu}\eqno ({\rm VII}.2)$$
where $d(\lambda )$ is the dimension of the representation $\lambda$.

Any complex-valued class function on $G$ can be expressed as a linear
combination of the characters:
$$\eqalignno{F(g)&=F(hgh^{-1})\hskip 1cm\Rightarrow\cr F(g)&=\sum_{\lambda}
A_{\lambda}^{(G,F)}X_{\lambda}(g)&({\rm VII}.3)\cr\noalign{\hbox{and}}
A_{\lambda}^{(G,F)}&=\int dg\; F(g)X_{\lambda}(g^{-1})&({\rm VII}.4)\cr
\noalign{\hbox{Specially}}\exp\big[ xF(g)\big]&=\sum_{\lambda}I_{\lambda}
^{(G,F)}(x)X_{\lambda}(g)&({\rm VII}.5)\cr I_{\lambda}^{(G,F)}(x)&=
\int dg\exp \big[ xF(g)\big] X_{\lambda}(g^{-1})&({\rm VII}.6)\cr}$$

\noindent{\lbf Acknowledgement}

\noindent I would like to express my deep gratitude to prof. R. Mansouri
for very useful discussions and encouragement.
\vfil\break

\noindent{\lbf References}

\item{[1]} K. G. Wilson, Phys. Rev. {\bf D10}, No. 8, P 2445, (1974)

\item{[2]} F. J. Wegner, J. Math. Phys. {\bf 12}, No. 10, P 2559, (1971)

\item{[3]} J. B. Kogut, Rev. Mod. Phys. {\bf 51}, No. 4, P 659, (1979)

\item{[4]} R. Balian et. al., Phys. Rev. {\bf D10}, No. 10, P 3376, (1974)

\item{[5]} C. Rebbi ed., ``lattice Gauge Theories and Monte Carlo
Simulations'',
 chapter 2, World Scientific publisher

\item{[6]} R. Balian et. al., Phys. Rev. {\bf D11}, No. 8, P 2098, (1975)

\item{[7]} M. Caselle et. al., DUFT 38/92,July 92, preprint